\documentstyle [12pt] {article}

\begin{document}



\newcount\itemnum  
\def\newitem{\itemnum=0}
\def\nextitemsup{\advance\itemnum by1 \item{$^{\the\itemnum}$}}
\def\nextitemp{\advance\itemnum by1 \item{{\the\itemnum}.}}

\newcount\refnum 
\def\newref{\refnum=0}
\newref

\newcount\eqnum 
\def\neweq{\eqnum=0}
\neweq

\newcount\istyle 
\def\stylepr{\istyle=1} 
\def\stylenp{\istyle=2} 
\def\styleap{\istyle=3} 

\newcount\localref 
\def\newloc{\localref=0}
\newloc

\def\nextref#1{\advance\refnum by1
    \header\outref{\the\refnum}\trailer
    \edef#1{\the\refnum}}
\def\reff#1{\header\outref{#1}\trailer} 
\def\nextrefs#1{\advance\refnum by1 \header
    \outref{\the\refnum,\hskip -.41em plus .08em}\edef#1{\the\refnum}
    \advance\localref by1}
\def\refs#1{\header \outref{#1,\hskip -.41em plus .08em}
    \advance\localref by1}
\def\outref#1{\ifnum\istyle=3{#1}\else{$^{#1}$}\fi}
\def\header{\ifnum\localref=0{\ifnum\istyle=3{[}\fi}\fi}
\def\trailer{\newloc \ifnum\istyle=2{$^)$}\fi \ifnum\istyle=3{]}\fi}

\newcount\secnum 
\def\newsec{\secnum=0}
\newsec

\newcount\subsecnum 
\def\newsubsec{\subsecnum=0}
\newsubsec

\def\section#1{\advance\secnum by1 \outsec{\the\secnum}{#1}
\newsubsec
     \ifnum\istyle=2\neweq\fi}
\def\outsec#1#2{\ifnum\istyle=2 \centerline{#1. #2}
     \else \centerline{\uppercase\expandafter{\romannumeral #1}.
     \uppercase{#2}} \fi}

\newcount\locoffset 

\def\numlet#1{\locoffset=96 \advance\locoffset by #1
     \char\locoffset}
\def\Numlet#1{\locoffset=64 \advance\locoffset by #1
     \char\locoffset}

\def\subsection#1{\advance\subsecnum by1 \vskip 0.6truein
\outsubsec{#1}}
\def\outsubsec#1{\ifnum\istyle=2
\leftline{\the\secnum.\the\subsecnum\ \  #1}
     \else\centerline{\Numlet{\subsecnum}.\  #1}\fi}

\def\pagenumbers{\footline={\hss\tenrm\folio\hss}}

\def\nexteq{\global\advance\eqnum by1
     \ifnum\istyle=
2{(\the\secnum}.{\the\eqnum})\else{(\the\eqnum})\fi}

\newcount\subeqlet 
\def\nexteqp{\global\advance\eqnum by1 \global\subeqlet=1 \outeqp}
\def\sameeq{\global\advance\subeqlet by1 \outeqp}
\def\outeqp{\ifnum
     \istyle=2{(\the\secnum}.{\the\eqnum}\numlet{\subeqlet})%
     \else{(\the\eqnum}\numlet{\subeqlet})\fi}

\stylenp        


\newcommand {\ket} [1] { | \, #1  > }

\newcommand {\bra} [1] { <  #1 \, | }

\newcommand {\ph} [1] { (-1)^{#1} }

\newcommand {\cgc} [3]
  { \mbox{ {\small\bf $ ( #1 : #2 |  #3 ) $ } }  }

\newcommand {\cgcsimp} [2] { \cgc{p_{#1},q_{#1};k_{#1},l_{#1},m_{#1}}
{p_{#2},q_{#2};k_{#2},l_{#2},m_{#2}} {p,q;k,l,m} }

\begin{center}
{\large\bf  Symmetry properties of SU3 vector coupling coefficients}

\bigskip
{\it H. T. Williams} \\
\smallskip
Department of Physics, Washington and Lee University, \\
Lexington, VA 24450, USA \\
\bigskip

\medskip\medskip
{\bf Abstract}
\medskip
\end{center}

{\small
A presentation of the problem of calculating the vector coupling coefficients
for $SU3 \supset SU2 \otimes U1$ is made, in the spirit of traditional
treatments of SU2 coupling.  The
coefficients are defined as the overlap matrix element between product states
and a coupled state with good SU3 quantum numbers.  A technique for resolution
of the outer degeneracy problem, based upon actions of the infinitesimal
generators
of SU3 is developed, which automatically produces vector coupling coefficients
with symmetries under exchange of state labels which parallel the familiar
symmetries of the SU2 case. An algorithm for efficient computation of these
coefficients is outlined, for which an ANSI C code is available.
}
\clearpage

{\flushleft {\bf I. Introduction}}
\medskip

The algebra associated with the irreducible representations (hereafter called
"irreps") of the special unitary group in two dimensions, SU2, has
proven itself an invaluable tool in
describing the angular momentum properties of quantum states
\nextrefs{\biedamt} \nextrefs{\edmamt} \nextref{\roseamt}.  In an early
manifestation of group symmetry as an elementary particle classification
scheme,
SU2 was recognized also to be the appropriate group for formalization of the
implications
of charge invariance through the concept of isospin.  The vector coupling
coefficients, \nextref{\vccs}
3-j, 6-j, and 9-j symbols, and recoupling techniques have become part
of the standard arsenal for dealing with quantum amplitudes in particle,
nuclear.
atomic, and molecular physics.

The group SU3 was recognized in the late 1950's
to be reflective of the symmetries inherent in the properties of the known
"elementary" hadrons \nextref{\gm-nm}.  Nearly simultaneously, its utility in
classification
of rotational states in non-spherical nuclei (due to the fact that the three
dimensional harmonic oscillator potential has SU3 symmetry) was exercised
\nextref{\elliott}.
The introduction of the color quantum number to explain the absence of
"exotic" hadronic states in the quark model, leading to Quantum Chromodynamics
which has proven successful in explaining the dynamics as well as the mass
spectra of elementary particles, has made further use of SU3 as the color
symmetry
of QCD.  Despite the subsequent discovery of additional quark flavors,
due to the extreme symmetry breaking brought on by the heavy masses
of the top and bottom quarks, it is still often practical to use SU3 flavor
symmetry
in calculations based upon quark structure of hadrons.

While the coupling and recoupling algebra of SU2 has been long known, nearly
standardized, and universally utilized, that for SU3 (despite its long history)
is much more poorly known, and is often underutilized in problems where it is
potentially
quite useful.  The reasons are several.  The algebra, as one would expect,
is more complicated than that for SU2.
The irreducible representations are labeled by two
integers (the set $p,q$
analogous to $j$ for SU2), and the basis states within an irrep are labeled
by three integers \nextref{\decomp}
($k,l,m$ with $p+q \geq k \geq q$, $q \geq l \geq 0$, and
$k \geq m \geq l$ in place of $j \geq m \geq -j$ for SU2).
Most of the
work done on the SU3 case has been presented a highly abstract language and
notation
rather unfamiliar to most physicists, and it has
often been published in journals not typically seen by the majority of
practitioners.  The problem known as "outer multiplicity" does not occur in
the SU2 case, further complicates the SU3 case, and inserts another point
in the algebra where some convention must be chosen to present unambigous
results.  As is typically the case, this has lead to several competing
conventions,
each of which has been created for the particular convenience or world view
of the author.  This presentation is intended to introduce a set of conventions
which will be of greatest convenience to those wishing to use SU3 as a tool
in the understanding of quantum state classification.

The notation used herein has been chosen to reflect the similarities of the
SU3 problem to the familiar SU2 classification of angular momentum states.  For
this reason, for example, a fully specified SU3 state will be denoted by
a ket, $\ket{p,q;k,l,m}$ in analogy to the angular momentum state $\ket{j;m}$
rather than the Gel'fand-Weyl notation frequently found in the mathematical
physics literature
\[ \left| \begin{array}{ccccc}
	p+q &   & q &   & 0 \\
	       & k &   & l  & \\
		 &  & m & & \\
	\end{array} \right> .  \]
The variables $p$ and $q$ distinguish the various irreducible representations
of SU3, and will here be referred to as {\em irrep labels}; $k$, $l$, and $m$
enumerate the basis states within the vector space acted upon by the irrep, and
will be referred to as {\em subspace labels}.
In elementary particle classification of states through flavor SU3, the
subspace labels $k,l,m$ are related to the quantum numbers
for total isospin ($I$), its projection ($I_z$), and hypercharge
($Y$), as follows:
\begin{eqnarray}
	I & = & \frac12 (k-l) \\
	I_z & = & m - \frac12(k+l) \\
	Y & = & k+l-\frac23 (p+2q) .
\end{eqnarray}
In the use of SU3 to classify nuclear rotational states, a variable $\Lambda$
and
its projection $M_\Lambda$ play roles analagous to $I$ and $I_z$, and
the variable $\epsilon$ corresponds to $-3Y$.  Letting $n_i$ represent
the number of harmonic oscillator quanta in the $i$ direction,
\begin{eqnarray}
	M_\Lambda = (n_1-n_2)/2, \\
	\epsilon = 2 n_3 - n_1- n_2.
\end{eqnarray}
When it helps to clarify meaning, the elementary particle notation will be used
herein,
and states will appear as $\ket{p,q;I,I_z,Y}$

\medskip\medskip
{\flushleft {\bf II. Description of the problem}}
\medskip

At the heart of the coupling/recoupling problem for a particular symmetry
group is the definition of the vector coupling coefficient (for SU3, this
will be at times hereafter be abbreviated VCC).  Considering the
direct product of two states, each classified according to SU3 quantum numbers,
it will be possible to express this as a weighted sum of SU3 classified states
of the composite system;
\begin{equation}
\ket{p_1,q_1;k_1,l_1,m_1} \, \ket{p_2,q_2;k_2,l_2,m_2} = \sum_i
 C_i \, \ket{p,q;k,l,m}.
\end{equation}

The weights in this sum are known as the vector coupling coefficients -- they
depend upon all fifteen variables describing the
three states involved, and will be denoted by
\[ C_i = \cgcsimp{1}{2}. \]
A choice of phases for the SU3 states can be made which will insure that the
vector coupling coefficients are real.  This fact, and the orthonormality
of the state kets allows the above relation to be inverted, so as to express
a composite SU3 state as a sum over product states, as
\begin{eqnarray}
\ket{p,q;k,l,m} & = & \sum_{k_1,l_1,m_1,k_2,l_2,m_2} \cgcsimp{1}{2} \nonumber
\\
  & & \ket{p_1,q_1;k_1,l_1,m_1} \, \ket{p_2,q_2;k_2,l_2,m_2}. \label{CGCdef}
\end{eqnarray}

Operators which map the space of these ket states into itself can also be
classified by their transformation properties under the group SU3.  It can
be shown that the matrix elements of such operators between SU3 states
is proportional to the vector coupling coefficients.  This relationship is
known as the Wigner-Eckart Theorem:
\begin{eqnarray}
 & \bra{p_f,q_f;k_f,l_f,m_f} T_{p,q;k,l,m} \ket{p_i,q_i;k_i,l_i,m_i} =
     \nonumber \\
     & \cgc{p,q;k,l,m}{p_i,q_i;k_i,l_i,m_i}{p_f,q_f;k_f,l_f,m_f}
     < \, p_f,q_f \, \| T_{p,q} \| \, p_i,q_i \, >,
\end{eqnarray}
where $T$ is the SU3 classified operator, and
\[ < \, p,q \, \| T_{p_1,q_1} \| \, p_2,q_2 \, >,  \]
the "reduced matrix element" is a complex number which depends only upon the
irrep lables.  It is possible, therefore, to define the VCC's
as matrix elements of a set of unit tensor operators.  Determination of a
set of unit tensor operators is then tantamount to the determination of the
vector coupling coefficients.

The requirement in equation \ref{CGCdef} that all kets be eigenstates of the
appropriate
group operators imposes severe restrictions on the vector coupling
coefficients.
The subspace labels must satisfy relations usefully thought of as
\begin{description}
	\item 1) conservation of hypercharge ($Y_1+Y_2 = Y$)
\begin{equation}
k_1+l_1 - \frac23(p_1+2q_1) + k_2+l_2-\frac23(p_2+2q_2) =
   k+l-\frac23(p+2q); \label{Y}
\end{equation}
	\item 2) conservation of charge ($I_{1z} +I_{2z} = I_z$)
\begin{equation}
m_1 - \frac12(k_1+l_1)+m_2-\frac12(k_2+l_2) = m-\frac12(k+l); \label{Iz}
\end{equation}
	\item3) triangularity relation for isospin ($\vec{I_1}+\vec{I_2}=\vec{I}$)
\begin{eqnarray}
	k_1-l_1+k_2-l_2 & \geq & k-l, \nonumber \\
	|k_1-l_1-k_2+l_2| & \leq & k - l.   \label{I}
\end{eqnarray}
\end{description}
As well, the "betweeness" relations
\begin{equation}
	p+q \geq k \geq q \mbox{  ,  } q \geq l \geq 0 \mbox{  , and  }
	k \geq m \geq l      \label{between}
\end{equation}
must be satisfied by each set of $k,l,m$'s.

There are, as well, restrictions on which sets of irrep labels correspond to
non-vanishing
vector coupling coefficients, in analogy to the triangularity condition for
angular momentum vectors for SU2, but rather more complicated.  The Clebsch
Gordan series is a formal expression of which irrep products contribute
to a particular composite irrep.  Letting $(p,q)$ represent the irreducible
representation labeled by indices $p$ and $q$, and $\alpha$ represent the
number of times a particular composite irrep appears in the outer product
of irreps $(p_1,q_1)$ and $(p_2,q_2)$,
\begin{equation} (p_1,q_1) \otimes (p_2,q_2) =
\sum_{p,q} \alpha(p_1,q_1,p_2,q_2,p,q) \, (p,q), \label{multiplicity}
\end{equation}
where
\[ \alpha = \max(\alpha '(p_1,q_1,p_2,q_2,p,q),0),  \]
and
\begin{eqnarray*}
   & \alpha '(p_1,q_1, p_2,q_2,p,q) = 1 + \\
   & min(q_1,q_2,p,p_1+\sigma - \gamma ,p_2+\sigma-\gamma,q+\sigma-\gamma,
   p_1+q_1-2\gamma-\sigma, p_2+q_2-2\gamma-\sigma, \gamma + 2 \sigma),
\end{eqnarray*}
   with
\begin{eqnarray*}
	\gamma & = & \frac13 \,\max(p_1+p_2-p,q_1+q_2-q), \\
	\sigma & = & \frac13 \,\min(p_1+p_2-p,q_1+q_2-q).
\end{eqnarray*}
The multiplicity $\alpha$ is  equal zero for "forbidden"
couplings for which all associated VCC's vanish identically, and is
a positive integer in all remaining cases, signifying the possibility of
non-vanishing VCC's (depending upon the values of the
subspace labels).  The above expression for the multiplicity
results from the betweeness conditions (equation \ref{between}) for the
subspace labels,
and the conservation laws required by the group symmetry (equations
\ref{Y}, \ref{Iz}, \ref{I}).  Other
equivalent expressions for this quantity, less useful in the present context,
appear in the literature \nextref{\intertwine}.

In the case of SU2 vector coupling coefficients, the multiplicity is either
one or zero.  In contrast, the multiplicity in the SU3 case can be arbitrarily
large.  The existence of multiplicities greater than one is the circumstance
typically referred to as the outer multiplicity problem.  For a
coupling of degeneracy $\alpha$,
the implication is as follows:  for two quantum states
put in combination leading to a SU3 classified composite state
$\ket{p,q;k,l,m}$ there are $\alpha$ such composite states which are distinct
in their physical properties specified by non-SU3 variables.
In effect, there is an $\alpha$ dimensional subspace of states with the same
SU3 description, and one must go beyond the requirements of SU3 symmetry to
chose $\alpha$ distinct states in this subspace to serve as a basis
for the degenerate subspace.  Only then are the vector coupling coefficients
determined (to within a sign.)  Here, the user must chose a convention, so that
the
resulting states are as free of arbitrary choice as possible, and which are
as useful as possible for his or her purposes.

\newpage
\medskip\medskip
{\flushleft {\bf III. Prior solutions to the problem}}
\medskip

An early published work on the SU3 vector coupling coefficients, which grew
from the
requirements of the quark model \nextref{\deswart}, gave a clear and
concise description
of the problem and related mathematical tools.  Values for a
limited set of VCC's are presented there in tables, including
cases of multiplicity two.  One such case was appropriate to the
problem of pion-nucleon scattering -- the coupling of an 8 dimensional
representation ($p=1$,$q=1$) to another 8.  The Clebsch Gordan series
for this product is
\[ (1,1) \otimes (1,1) = (0,0) \oplus 2 \, (1,1) \oplus (3,0) \oplus (0,3)
\oplus (2,2). \]
The twofold appearance of $(1,1)$ in the product is the simplest case in
SU3 where outer multiplicity appears, but evaluation of VCC's
for this case does not require a general resolution of this issue:  a simple
parity
condition (coefficients  either even or odd with respect to interchange
of particles one and two) suffices to define the coefficients within a sign.
The same set of coefficients for this case result from several quite distinct
choices
of resolution criteria, which produce very different results in other cases.

Formal work on the general problem of outer multiplicity was undertaken
\nextrefs{\bieda} \nextref{\biedb} and the problem was
formally resolved as was exhibited in a 1970 paper \nextref{\biedc} which
derived results
for general matrix elements of the $(p,q)=(2,2)$ tensor operator, which
admits one case of threefold multiplicity.  Further development and
clarification
of this solution has continued \nextrefs{\biedd} \nextrefs{\biede}
\nextref{\alis}.

A simplified summary of the agenda of Biedenharn, Hecht, Louck and
collaborators for resolution of the outer degeneracy problem is
illustrative.  This agenda will henceforth be referred to as the {\em
canonical}
resolution, adopting the language of those authors.
The focus is on the determination of unit tensor operators
which can be classified according to the SU3 quantum numbers $p,q;k,l,m$
according to their behavior under the action of the group generators.  The
general nature of the Clebsch Gordan series shows that such an operator has
non-vanishing matrix elements between states from several different irreps.
Further restricion of the operators is made to those which produce a specific
set of shifts -- changes in the irrep lables of a state $(p,q)$ resulting from
the action of
the operator thereon.  In cases of non-trivial outer degeneracy, even these
restricted operators produce ambiguous results, since there is more than one
version of a particular set of irrep variables which results from the
application
of the operator on a state $(p,q)$.  The complete resolution of the problem is
to define in these
cases a unique set of operators with the same SU3 labels which produce the same
shifts, and which produce a unique single version of the shifted state.

It was shown that such a completely restricted tensor must be labled by
eight integers:  $p,q,k,l,m$ representing the SU3 transformation
properties of the operator, and a further set
$\kappa,\lambda,\mu$ which properly imply the restrictions required.  This
triplet
of integers, called the operator pattern, satisfies betweeness conditions
like those for the $k,l,m$:  $p+q \geq \kappa$, $q \geq \lambda \geq 0$,
$\kappa \geq \mu \geq \lambda.$  The shifts,
\begin{eqnarray*}
	\Delta p = 2 \mu - \kappa - \lambda  \\
	\Delta q = 2 \kappa + 2 \lambda - \mu - 2p - q
\end{eqnarray*}
indicate the effect of the opertor on the irrep indices of the state
upon which it operates:  in terms of the operator's reduced matrix element,
\[ < \, p_f,q_f \, \| T_{p,q;k,l,m;\kappa,\lambda,\mu} \| \, p_i,q_i \, > = 0
\]
except in cases when
\[ p_f-p_i = \Delta p \; \; \mbox{and} \; \; q_f - q_i = \Delta q . \]
The remaining freedom in the operator pattern lables, that is the
range of values of
$\kappa$ and $\lambda$ which obey betweeness and have a fixed sum (thus
producing always the same shifts) precisely lables the degeneracy of the
coupling of $(p,q)$ to $(p_i,q_i)$ to produce $(p_f,q_f)$.

Distinctions among operators having the same SU3 labels and shifts
is made using the concept of a {\em characterisic
null space}.  This refers to the union of all irreps $(p_i,q_i)$
which when operated upon
by the tensor operator $T$ yield a zero result.  In the case of a tensor
operator
without degeneracy ($\alpha = 1$), the null space is precisely determined by
the restrictions the group imposes upon the operator and the state upon
which it operates.  The only other operator which has the
same group labels and shifts, and a different null space than the previous one
is the null operator.  For degenerate operators , this is not the case,
and the various operators of a given set of group lables and shifts can be
identified with a set of operators with null spaces, each larger than
the previous and completely containing it.  Construction of operators with
precisely these null space properties was accomplished through a build-up
process using SU3 operators without dengeneracy, whose matrix elements had
algebraic expressions \nextref{\biedf}.  Continuing work
to make explicit the operators and vector coupling coefficients so generated
has produced a mapping of the tensor operators onto an SU3 invariant norm,
called the denominator function, for which explicit expressions involving
ratios of polynomials have been derived \refs{\biedd} \refs{\biede}
\reff{\alis}

A complete implementation of the canonical operator build-up scheme
exists in the form of computer codes generated by  Draayer
et.al. \nextrefs{\draaa} \nextrefs{\draab} \nextrefs{\draac}
\nextref{\draad}.  They exhibit that each tensor operator of
multiplicity greater than one can be built up from an operator of multiplicity
one
(whose matrix elements are simply computable) by multiple products with a
special
set of tensor operators of $(p,q) = (1,1)$ with shifts $\Delta p = \Delta q =
0$
which are the infinitesimal generators of SU3.  Their results are shown to
generate precisely the canonical vector coupling coefficients of Biedenharn et.
al..  The current manifestation of this computer work is a 3600 line set
of Fortran codes which computes SU3 vector coupling coefficients (both for the
SU2xU1 decomposition and the R3 decomposition) as well as 6-j and 9-j symbols
constructed therefrom.

An alternative scheme for outer multiplicity, but similar in spirit to the
canonical
scheme just described, was recently presented \nextref{\lbr}.
It, too, relates the VCC's to matrix elements of tensor operators, and relies
heavily
upon prior work on the so-called Bargmann tensors.  It is shown how known
Bargmann
tensors can be used to build the desired tensor operators for SU3, from which
the
VCC's are easily computed.  A direct (non-recursive) way of computing matrix
elements of the Bargmann tensors of interest is given, and from that, the
VCC's are produced using a Gramm-Schmidt orthogonalization procedure.  The
tensors here differ from those of the canonical scheme in their null space
properties, and as a consequence, the Wigner-Eckart theorem involves a sum
over the outer multiplicity label.  This necessitates the Gram-Schmidt
procedure
to extract the VCC's from the matrix elements of the unit tensor operators.
The
ordering of the tensors which is necessary to specify the orthogonalizaton
algorithm is based upon a conjecture due to Braunschweig \nextref{\braun}
which is straightforwardly verified in the work presented below.

\medskip\medskip
{\indent {\bf IIIa. Symmetry properties of coefficients}}
\medskip

The physical process which corresponds to the coupling of a particular SU3
classified quantum state to another is the observation of the two particles
in combination, either as a bound state or an intermediate state of a
scattering
event or reaction.  The probability of creation of a particular such state
is proportional to the square
of the associated vector coupling coefficient.  The order of coupling (which
physical state is associated with state "1" and which with state "2")
is not observable, so in order that the coupled state have an interpretation
as a possible physical state, it is necessary that the vector coupling
coefficients $\cgcsimp{1}{2}$ and $\cgcsimp{2}{1}$ differ by, at most, a sign.
A similar symmetry regarding the interchange of the first and last set of
indices would allow the definition of a proportional coefficient with
at most a sign change under interchange of any of the three sets of SU3
quantum numbers -- the 3-j symbol.  This forms the basis for straightforward
definitions of 6-j and 9-j recoupling symbols with high degrees of symmetry
under interchange of indices.
The SU2 vector coupling coefficients have these symmetries.  As well, all SU3
VCC's with multiplicity
one have them as well.  When a method of resolution of the outer multiplicity
degeneracy is prescribed, there is no guarantee that these desirable properties
will be possessed by the resulting VCC's.  The coefficients of the canonical
resolution, as calculated by the Draayer codes, do not have these interchange
symmetries.
Because the alternative resolution of Le Blanc and Rowe, like the canonical
one,
resolves the degeneracy by consideration of properties of the operator (in
the Wigner-Eckart theorem) which places it in an unsymmetrical relationship
with the initial state tensor properties, it seems quite likely that the
exchange symmetries are also missing in those results.
The existence of VCC's with such symmetries for arbitrary groups has been
investigated
\nextrefs{\exista} \nextrefs{\existb} \nextref{\existc}
and it was shown that while it was not, in general, possible (it is
impossible, for example, for S6 -- the symmetric group on six symbols), in
the particular case of SU3, such VCC's indeed existed.
A scheme for the generation of SU3 VCC's with interchange symmetry was devised
by Pluhar, et. al. in 1986 \nextref{\pluhar}.  These authors create a labelling
operator
constructed from representation generators and show that finding basis states
within the degenerate subspace of a multiplicity $>1$ coupling, which were
eigenstates of the labelling operator, would guarantee interchange symmetry.
The
eigenvalues of the labelling operator become the missing label.  The lables,
in general, are irrational numbers.
This implies that the VCC's which result cannot be represented as the square
root of ratios of integers, a useful property familiar from the SU2 case, and
which has been proven to be the case for the coefficients of the canonical
scheme \reff{\biedd}.

\medskip\medskip
{\flushleft {\bf IV. Symmetric resolution of outer degeneracy}}
\medskip

For many theoretical and experimental
physicists, the coupling and recoupling Racah algebra for SU2 is a tool
required for even routine calculations.  Those parts of the algebra necessary
for this kind of calculation are typically learned in courses in graduate
school from advanced quantum mechanics texts like Messiah \nextref{\messiah},
or
specialized monographs like Edmunds \reff{\edmamt}, or Rose \reff{\roseamt}.
There, the
vector coupling coefficients are introduced as elements of a transformation
matrix from a set of basis states (of a two particle system) which are product
of single particle states,
\[ \ket{j_1;m_1} \, \ket{j_2;m_2}  \]
each simultaneously an eigenvector of the angular momentum operator and
the z-component of angular momentum operator; to a set of states
\[ \ket{ j_1,j_2,J;M} \]
which is an eigenvector of the total (two particle) angular momentum operator
and its corresponding z-component operator.  The connection of the VCC's
to matrix elements of tensor operators is presented after the coefficients
are fully explicated, through the Wigner-Eckart Theorem.  Properties of
the VCC's and explicit formulae for their evaluation are achieved by means
of $J_+$ and $J_-$, the angular momentum raising and lowering operators,
two of the infinitesimal generators of the group SU2.

The motivation for the present work is twofold:  to present the SU3 Clebsch
Gordan problem in terms as similar as possible to this familiar agenda from
the SU2 problem, and to produce in this language a straightforward means
of resolving the outer degeneracy problem which will preserve certain desirable
and familar properties of symmetry under interchange of particle lables.

For a system whose Hamiltonian is invariant under the operations of the SU3
group,
the degeneracy structure of its energy spectrum is labeled by the irreducible
representations of the group.  Allowed energy eigenvalues can be labled by two
integers $p,q$ which are related to the eigenvalues of the Casimir operators
of the group.  Each energy $E_{p,q}$ is shared by exactly $d = \frac12 (p+1)
(q+1)(p+q+2)$ distinct eigenvectors, where $d$ is the dimension of the
irrep $(p,q)$.  An orthonormal set of basis states for the irrep can be
chosen by picking states which are simultaneously eigenvalues of the three
commuting operators ${\hat T}_3$, $\hat Y$, and ${\hat T}^2 = \frac12(\hat{T}_
+ {\hat T}_- + {\hat T}_- {\hat T}_+)-{\hat T}_3^2.$
The operators ${\hat T}_+,{\hat T}_-,{\hat T}_3,\hat Y$ are four of eight
infinitesimal generators
of the group, and are defined by their mutual commutation relations.  The
relationships among these generators and their relation to the canonical
generators can be found in a classic elementary presentation of
the theory from the point of view of elementary particle classification, by
Gasiorowicz and Glashow \nextref{\gasgla}, whose notation is
adopted here.  The states of the orthonormal basis
for the irrep $(p,q)$ are defined by the eigenvalue equations:
\begin{eqnarray}
	{\hat T}_3 \ket{p,q;k,l,m} & = & (m - \frac12 (k+l)) \ket{p,q;k,l,m}  =
	I_z \ket{p,q;I,I_z,Y} \\
	\hat Y \ket{p,q;k,l,m} & = & (k+l-\frac23 (p+2q)) \ket{p,q;k,l,m}  =
	Y \ket{p,q;I,I_z,Y} \\
	{\hat T}^2 \ket{p,q;k,l,m} & = & \frac14(k-l)(k-l+2) \ket{p,q;k,l,m}  =
	I(I+1) \ket{p,q;I,I_z,Y}.
\end{eqnarray}
The remaining four infinitesimal generators, along with two already mentioned
$({\hat T}_+,{\hat T}_-)$, transform among states within an irrep, playing a
role akin to the angular momentum raising and lowering operators:
\begin{eqnarray}
	{\hat T}_+ \, | \, p,q; \, k,l,m \, > & = & \sqrt{(k-m)(m-l+1)} \,
	       | \, p,q; \, k,l,m+1 \, > \label{tplus} \\
	{\hat T}_- \, | \, p,q; \, k,l,m \, > & = & \sqrt{(k-m+1)(m-l)} \,
	       | \, p,q; \, k,l,m-1 \, > \label{tminus} \\
	{\hat V}_+ \, | \, p,q; \, k,l,m \, > & = &
	     \sqrt{\frac{(k+2)(m-l+1)(k-q+1)(p+q-k)}{(k-l+1)(k-l+2)}} \,
	       | \, p,q; \, k+1,l,m+1 \, > \nonumber \\
	 & & + \sqrt{\frac{(l+1)(k-m)(q-l)(p+q-l+1)}{(k-l)(k-l+1)}} \,
	       | \, p,q; \, k,l+1,m+1 \, > \label{vplus} \\
	{\hat V}_- \, | \, p,q; \, k,l,m \, > & = &
	     \sqrt{\frac{(k+1)(m-l)(k-q)(p+q-k+1)}{(k-l)(k-l+1)}} \,
	       | \, p,q; \, k-1,l,m-1 \, > \nonumber \\
	 & & + \sqrt{\frac{l(k-m+1)(q-l+1)(p+q-l+2)}{(k-l+1)(k-l+2)}} \,
	       | \, p,q; \, k,l-1,m-1 \, > \label{vminus} \\
	{\hat U}_+ \, | \, p,q; \, k,l,m \, > & = &
	     \sqrt{\frac{(k+2)(k-m+1)(k-q+1)(p+q-k)}{(k-l+1)(k-l+2)}} \,
	       | \, p,q; \, k+1,l,m \, > \nonumber \\
	 & & - \sqrt{\frac{(m-l)(l+1)(q-l)(p+q-l+1)}{(k-l)(k-l+1)}} \,
	       | \, p,q; \, k,l+1,m \, > \label{uplus} \\
	{\hat U}_- \, | \, p,q; \, k,l,m \, > & = &
	     \sqrt{\frac{(k+1)(k-m)(k-q)(p+q-k+1)}{(k-l)(k-l+1)}} \,
	       | \, p,q; \, k-1,l,m \, > \nonumber \\
	 & & - \sqrt{\frac{l(m-l+1)(q-l+1)(p+q-l+2)}{(k-l+1)(k-l+2)}} \,
	       | \, p,q; \, k,l-1,m \, > \label{uminus}.
\end{eqnarray}
The operators ${\hat T}_+$ and ${\hat T}_-$ move up and down in the
variable $I_z$ at constant value of $I$ and $Y$.  The remaining
operators move simultaenously to neighboring states in the $I$, $I_z$, and
$Y$ grid in such a way that appropriate products and sums of these
six operators allow one to generate any state in a multiplet (of constant
$p,q$) from any other in the same multiplet.

These operators simplify the
outer degeneracy resolution problem, in that if one can find a means to
produce a single state within a set of irrep basis states, properly resolved
and labeled according to outer multiplicty, all other states within the
multiplet
(and thus all VCC's with the same value of $(p,q)$) follow immediately.
This advantage has been frequently utilized in the prior work cited above.
Here,
a precise algorithm for degeneracy resolution will be generated for the "state
of highest weight" (SHW) for a particular coupling, which is defined to be
the composite state with $k=m=p+q$, $l=0$.
This state has the largest values of $I$ and $I_z$ of all states in the
multiplet.
Knowledge of this state implies knowledge of all the VCC's of the form
\[ \cgc{p_{1},q_{1};k_{1},l_{1},m_{1}}
{p_{2},q_{2};k_{2},l_{2},m_{2}} {p,q;p+q,0,p+q}.  \]
The remaining VCC's for the same values of the irrep labels can be
deduced iteratively from relations generated by use of the raising and
lowering operators above.

In order to determine the product states which will be represented
in the composite SHW,
{\samepage \begin{eqnarray}
       & & \ket{p,q;p+q,0,p+q}  \equiv  \ket{p,q;SHW}  \\
& & \! = \! \sum_{k_1,l_1,m_1,k_2,l_2,m_2}  \! \!
   \cgcsimp{1}{2} \ket{p_1,q_1;k_1,l_1,m_1} \ket{p_2,q_2;k_2,l_2,m_2},
   \nonumber \label{defineSHW}
\end{eqnarray} }
necessary and sufficient conditions for product states with non-vanishing
VCC's come from the betweenness conditions (equation \ref{between})
for states $1$ and $2$,
and the conservation laws for the subspace labels, which read
for this state as
\begin{eqnarray*}
	k_1+l_1 - \frac{2}{3}(p_1+2q_1)+k_2+l_2-\frac{2}{3}(p_2+2q_2)  & = &
\frac{1}{3}(p-q) \\
	m_1- \frac{1}{2}(k_1+l_1) +m_2 - \frac12(k_2+l_2) & = & \frac12 (p+q) \\
	k_1-l_1+k_2-l_2 & \geq & p+q \\
	|k_1 - l_1 -k_2 +l_2 | & \leq & p+q .
\end{eqnarray*}

A set of conditions for the determination of the VCC's in this
sum (equation \ref{defineSHW}) come from the action of selected operators on
$\ket{p,q;SHW}$.
By inspection of this sum, which defines the state of highest weight, and
the properties of the raising and lowering operators {equations
\ref{tplus}, \ref{vplus}, and \ref{uminus}, it is
easily seen that
\begin{eqnarray}
	{\hat T}_+ \ket{SHW} & = & 0 \nonumber \\
	{\hat V}_+ \ket{SHW} & = & 0  {\mbox , and} \nonumber \\
	{\hat U}_- \ket{SHW} & = & 0.\label{opsonshw}
\end{eqnarray}

\medskip\medskip
{\indent {\bf IV.a. Multiplicity one}}
\medskip

For couplings of irreps which lead to a multiplicity of one (equation
\ref{multiplicity}), the
above conditions (equation} \ref{opsonshw}) are sufficient to determine all the
VCC's.  Each of the operators is linear, so that the operator for the
composite case is the sum of the operators for each of the single particle
states, for example
\begin{eqnarray}
& & {\hat V}_+ \ket{SHW}  =  ({\hat V}_{1+} + {\hat V}_{2+}) \ket{SHW}
\nonumber \\
& &  =  \sum_{k_1,l_1,m_1,k_2,l_2,m_2} \cgcsimp{1}{2}  \\
& &   \left( ({\hat V}_{1+}\ket{p_1,q_1;k_1,l_1,m_1}) \ket{p_2,q_2;k_2,l_2,m_2}
     + \ket{p_1,q_1;k_1,l_1,m_1} ({\hat V}_{2+}\ket{p_2,q_2;k_2,l_2,m_2})
\right)
     \nonumber \label{vplusonsum}.
\end{eqnarray}

The actions of the operators upon the single particle states are as in
equation \ref{vplus}.  The result is a sum over different product states, with
coefficients which are linear combinations of VCC's and coefficients
from equation \ref{vplus}.  Because product states with any difference in their
indices are orthogonal, the coefficient of each unique product state in
the sum (equation \ref{vplusonsum}) must vanish, producing a set of equations
which
must be satisfied by the VCC's.  The result may be represented by a
four term recursion relation for VCC's of various indices.  Similarly, the
action of
${\hat U}_-$ on the SHW produces a distinct four term recursion relation, and
that of ${\hat T}_+$ produces a two term recursion relation.  These
relationships
are not linearly independent, and the use of the ${\hat V}_+$ and ${\hat U}_-$
equations alone give maximal information.

The requirement that $\ket{p,q;SHW}$
be normalized, implies that the sum of squares of all the VCC's involved in
the summations of equation \ref{defineSHW} must be $1$.  Finally, the sign of
one of the
VCC's in equation \ref{defineSHW} must be chosen:  for reasons which will
become
clearer in the next section, the sign convention chosen is to require that
\begin{eqnarray}
	\cgc{p_{1},q_{1};p_1+q_1,0,p_1+q_1}
	{p_{2},q_{2};k_{2}^*,l_{2}^*,m_{2}^*} {p,q;p+q,0,p+q}  \equiv \nonumber \\
	\cgc{p_1,q_1;SHW} {p_{2},q_{2};k_{2}^*,l_{2}^*,m_{2}^*} {p,q;SHW} > 0,
\label{signconv}
\end{eqnarray}
where
\begin{eqnarray*}
 k_2^* & = & \frac13(p-p_1+2p_2-q+q_1+4q_2) \\
 l_2^* & = & 0 \\
 m_2^* & = & \frac13(2p-2p_1+p_2+q-q_1+2q_2),
\end{eqnarray*}
in cases where $p_1+p_2-p \geq q_1+q_2-q$; otherwise
\begin{eqnarray*}
 k_2^* & = & p_2+q_2 \\
 l_2^* & = & \frac13(q_1+q_2-q-p_1-p_2+p) \\
 m_2^* & = & \frac13(2p-2p_1+p_2+q-q_1+2q_2) .
\end{eqnarray*}

The combination of conditions represented by equations \ref{opsonshw},
the normalization condition, and the sign convention, completely determine
all the VCC's for any coupling of irreps with multiplicity one.  While
judicious handling of the recursion relations might yield an algebraic
closed form expression for any single VCC, the complexity of the resulting
formula would render it in most cases useless but for computer evaluation:
coding the equations and letting the computer carry out the required recursion
should be less error-prone, more comprehensible, and usually as efficient.

\newpage
\medskip\medskip
{\indent {\bf IV.b. Multiplicity greater than one}}
\medskip

When the values of the six irrep variables for a particular coupling produce
a value of the multiplicity ($\alpha$) which is greater than one, the combined
conditions of betweenness, equations \ref{opsonshw}, and normalization
do not totally determine the state
$\ket{SHW}$.  They instead give $(\alpha-1)$ fewer conditions than there
are unknown VCC's in the state $\ket{SHW}$.
The remaining conditions, necessary to resolve the
outer degeneracy, must come from outside the requirements of SU3 symmetry.
Here they are chosen to produce the symmetry under particle exchange properties
mentioned earlier, and to simplify computation of the VCC's.

A description of this outer multiplicity resolution technique will be
described which applies only to cases such that $p_1+p_2-p \geq q_1+q_2-q$.
For the
remaining cases, a technique which differs only in the details applies.
There will be presented later a symmetry property for the resulting VCC's
which will relate each VCC with $p_1+p_2-p < q_1+q_2-q$ to one with
$p_1+p_2-p > q_1+q_2-q$ with the same magnitude and a predictable sign, so
the restriction above is of little practical consequence.

It can be shown in a straightforward manner that each member of the following
series of VCC's does not violate any previously given conditions, and therefore
does not necessarily vanish:
\begin{equation}
\cgc{p_1,q_1;SHW} {p_{2},q_{2};k_{2}^*-j,j,m_{2}^*} {p,q;SHW} , \label{series}
\end{equation}
where, as before,
\begin{eqnarray*}
 k_2^* & = & \frac13(p-p_1+2p_2-q+q_1+4q_2) \\
 m_2^* & = & \frac13(2p-2p_1+p_2+q-q_1+2q_2).
\end{eqnarray*}
The integer variable $j$ ranges in steps of $1$ from $0$ to $j_{max}$, a
maximum variable limited only by the betweeness conditions for state 2,
which are
\begin{eqnarray*}
	k_2^* - j & \geq & m_2^* \\
	k_2^* - j & \geq & q_2 \\
	j & \geq & q_2 \\
	j & \geq & m_2^* .
\end{eqnarray*}
This leads simply to the expression
\[ j_{max} = \min(\gamma + 2 \sigma, p_2+ \sigma - \gamma, q_2, p_2+q_2 -
   2 \gamma -\sigma), \]
where, in this case, $\gamma = \frac13 (p_1+p_2-p)$ and $\sigma = \frac13
(q_1+q_2-q)$.  Inspection of equation \ref{multiplicity} for the degeneracy,
shows that the number of members of the series of equation \ref{series},
$j_{max}+1$, is at least as large as the multiplicity.  This proves what
has been referred to in the literature as Braunschweig's conjecture
\reff{\braun}, which has been important in several previous investigations
into the SU3 VCC's \reff{\lbr}.

Each member of the series given in equation \ref{series} can be seen as
representative of a family of all the VCC's for a particular coupling, with a
fixed
value of the {\em scalar} sum
\[ S \equiv I_1+I_2 = \frac12(k_1-l_1+k_2-l_2). \]
The largest value, $S_{max}$, corresponding to $j=0$, is
\[ S_{max} = \frac13 (p+2 p_1+2 p_2 - q +4 q_1 + 4 q_2); \]
for the $j$-th member of the series, the related family all have
\[ S = S_{max}-j . \]

The linear relations among the VCC's implied by the conditions of
the last two of equations
\ref{opsonshw} involve four VCC's -- two with $I_1+I_2 = S$ and two with
$S+1$.  For the family of VCC's with $S=S_{max}$, these become two-term
recursion relations which connect all the VCC's in the $S_{max}$ family.
Should one of the $S=S_{max}$ VCC's vanish, all will vanish; if so, then
the VCC's with $S=S_{max}-1$ will be connected by proportionalities for
the same reason.  Generalizing, if each member of the series (equation
\ref{series}) for $0 \leq j \leq j^*$ vanishes, then all VCC's with
$S \geq S_{max} - j^*$ will vanish:  as a result, all VCC's with
$S = S_{max}-j^*-1$ will be proportional to one another so that knowledge of
one fixes the values of all others.

To resolve the outer degeneracy, therefore, for a case of multiplicity
$\alpha$ ($ > 1$), follow the following steps:
\begin{description}
	\item 1) set
\begin{equation}
\cgc{p_1,q_1;SHW} {p_{2},q_{2};k_{2}^*-j,j,m_{2}^*} {p,q;SHW} = 0
\end{equation}
for $j=0,1, \ldots, \alpha-2$.
	\item 2) set
\begin{equation}
\cgc{p_1,q_1;SHW} {p_{2},q_{2};k_{2}^*-\alpha+1,\alpha-1,m_{2}^*} {p,q;SHW} =
c_1 > 1.
\end{equation}
This will allow direct calculation via recursion of all VCC's in the
$j = \alpha-1$ family, {\it and all remaining non-zero VCC's}, in
terms of the constant $c_1$.  After
chosing the value of the constant $c_1$ to satisfy the normalization
condition and sign convention for the VCC's,
a set of VCC's appropriate to the involved irrep labels is completely
determined.  These VCC's will carry the outer degeneracy
label $j = \alpha-1$. \\
	\item 3) to determine a second, orthogonal, set of VCC's,
with outer degeneracy label $j = \alpha - 2$, set
\begin{equation}
\cgc{p_1,q_1;SHW} {p_{2},q_{2};k_{2}^*-j,j,m_{2}^*} {p,q;SHW} = 0
\end{equation}
for $j=0,1, \ldots, \alpha-3$. \\
	\item 4)  set
\begin{equation}
\cgc{p_1,q_1;SHW} {p_{2},q_{2};k_{2}^*-\alpha+2,\alpha-2,m_{2}^*} {p,q;SHW} =
c_2,
\end{equation}
and
\begin{equation}
\cgc{p_1,q_1;SHW} {p_{2},q_{2};k_{2}^*-\alpha+1,\alpha-1,m_{2}^*} {p,q;SHW} =
c_1.
\end{equation}
The first condition allows calculation of all VCC's in the $j=\alpha-2$ family
in terms of the constant $c_2$, and the second allows a recursive evaluation
of all remaining non-zero VCC's in terms of $c_1$ and $c_2$.  The two constants
are determined by the condition of normalization and the requirement that
this set of VCC's must be orthogonal to the set previously determined, i.e.
\begin{eqnarray}
     && \sum_{k,l,m's} \cgc{p_1,q_1;k_1,l_1,m_1}{p_2,q_2;k_2,l_2,m_2}{p,q;k,l,m
\; [j]} \nonumber
      \\ && \; \cdot \cgc{p_1,q_1;k_1,l_1,m_1}{p_2,q_2;k_2,l_2,m_2}{p,q;k,l,m
\; [j']}
	= 0, \label{orthogonality}
\end{eqnarray}
for $j \neq j'$. \\
	\item 5) this process is repeated, each time setting one fewer of
the VCC's in the $j$ series equal zero, setting one more from this series
equal to a yet-undetermined constant, and ultimately determining all these
constants by the sign convention, normalization condition, and the
requirement that each of the various sets of $j$-labeled VCC's must be
orthogonal to all the remaining (equation \ref{orthogonality}).
The procedure is similar to the familiar
Gram-Schmidt process.

The sign convention chosen for each of the $j$-labeled sets of VCC's is
a generalization of that for the multiplicity one case;
\[ \cgc{p_1,q_1;SHW}{p_2,q_2;k_2^*-j,j,m_2^*}{p,q;SHW} > 0.  \]
\end{description}

The result of carrying this out to its conclusion is the determination
of exactly $\alpha$
distinct sets of VCC's which satisfy all the SU3 requirements for the
involved irreps, which correspond to mutually orthogonal and normalized
kets $\ket{p,q;SHW}$, and which satisfy a set of symmetry relations
which will be described in the next section.

\medskip\medskip
{\flushleft {\bf V. Symmetries of the Vector Coupling Coefficients}}
\medskip

The vector coupling coefficients of SU2 posess three symmetries, which
aid in their numerical evaluation, inspire the creation of the 3-j
symbol with its more natural symmetries, and greatly simplify the
evaluation and manipulation of the 6-j and 9-j recoupling coefficients.
In the case where angular momentum states $\ket{j_1;m_1}$ and $\ket{j_2;m_2}$
are coupled together to a good final angular momentum state
$\ket{j_1,j_2,j;m}$,
the symmetries are as follows:
\begin{description}
	\item 1) upon interchange of the values of $j_1,m_1$ with $j_2,m_2$,
	the VCC changes by, at most, a sign; \\
	\item 2) upon replacement of each $j_i,m_i$ pair with its conjugate,
	$j_i,-m_i$, the VCC changes by, at most, a sign; \\
	\item 3) upon interchange of the values of $j_1,m_1$ with $j,-m$,
	the VCC changes in magnitude by the square root of the ratio of
	the dimension of the multiplet $j_1$ to that of the multiplet
	$j$, i.e. $\sqrt{(2j_1+1)/(2j+1)}$; and can change in sign.
\end{description}
The exact expressions for the sign changes resulting from these alterations
depends on
the choice of sign convention, and varies from author to author.  The
symmetry under interchange of $j_2,m_2$ with $j,-m$ can be deduced from
the symmetries above.

The SU3 VCC's defined in the last section, including those with multiplicity
greater than one, can be shown to have a set of symmetries which entirely
parallel those for SU2.  Consider, for example, the interchange of the
set of indices $p_1,q_1,k_1,l_1,m_1$ with $p_2,q_2,k_2,l_2,m_2$, and the
steps involved in the evaluation of the VCC.  The product basis states
are invariant under the interchange.  The conditions which specify which
product states will correspond to non-vanishing VCC's -- the betweenness
conditions, and the conservation of $I,I_z$, and $Y$ -- are also invariant
under the interchange.  The conditions which completly specify the values
of the VCC's in the case of multiplicity one (equations \ref{opsonshw})
are also invariant, since each operator is the scalar sum of the operator
for state $1$, and that for state $2$.  In the case of higher multiplicity,
it is straightforward to show that the procedure is invariant, since
for every non-zero VCC
\[ \cgc{p_1,q_1;SHW}{p_2,q_2;k_2^*-j,j,m_2^*}{p,q;SHW} \]
there is a corresponding non-zero VCC
\[ \cgc{p_2,q_2;SHW}{p_1,q_1;k_1^*-j,j,m_1^*}{p,q;SHW} \]
which has the same value of $S = I_1+I_2$.  The VCC itself is not invariant
under interchange solely because the sign convention is not invariant.
The state which is chosen to be positive in the two cases ($(p_1,q_1) \otimes
(p_2,q_2) \rightarrow (p,q)$ and $(p_2,q_2) \otimes (p_1,q_1) \rightarrow
(p,q)$)
is in each case in the $S = S_{max}$ family.  VCC's in this family
are related by a two term recursion relation with alternating signs, and it
is straightforward to predict the sign change upon interchange:
\begin{eqnarray*}
 && \cgc{p_1,q_1;k_1,l_1,m_1}{p_2,q_2;k_2,l_2,m_2}{p,q;k,l,m [j]} \\
 && \; \; \; = \ph{2 \gamma + \sigma +j} \cgc{p_2,q_2;k_2,l_2,m_2}
 {p_1,q_1;k_1,l_1,m_1}{p,q;k,l,m [j]}
\end{eqnarray*}
where j is the degeneracy index, and $\gamma$,$\sigma$ are the maximum and
minimum, respectively, of $\frac13(p_1+p_2-p)$ and $\frac13(q_1+q_2-q)$.

The Hermitian conjugation operation, which maps the state $\ket{j,m}$ into
$\ket{j,-m}$
in SU2, maps $\ket{p,q;k,l,m}$ into $\ket{q,p;\tilde{k},\tilde{l},\tilde{m}}$
(where $\tilde{k} = p+q-l$, $\tilde{l}=p+q-k$, $\tilde{m}=p+q-m$) in SU3,
changing
not only the subgroup indices, but the irrep indices as well.  Otherwise
stated,
if an operator $T$ transforms like an SU3 tensor of indices $p,q;k,l,m$, then
its Hermitian conjugate, $T^{\dagger}$, transforms like one of indices
$q,p;\tilde{k},\tilde{l},\tilde{m}$.

If all three sets of SU3 indices in a VCC are simultaneously conjugated,
the numerical value of the VCC changes by, at most, a sign.  To establish
the phase in this relationship, it is necessary to examine
the VCC's for cases such that $q_1+q_2-q > p_1+p_2-p$, since simultaneous
conjugation of all states transforms a VCC with $p_1+p_2-p > q_1+q_2-q$
into one with the opposite inequality.  The steps described in the last
section for determining the VCC's and resolving outer degeneracy (if
non-trivial) are followed as well in cases such that $q_1+q_2-q > p_1+p_2-p$,
with one major change:  the series of VCC's which characterize the
families of constant $I_1+I_2$ and which fix the sign convention, consists of
\[ \cgc{p_1,q_1;SHW}{p_2,q_2;p_2+q_2-j,l_2^*+j,m_2^*}{p,q;SHW}. \]
It is easily seen that under conjugation, the various terms in the
infinitesimal generators transform among one another in a way that enables
a straightforward albeit tedious derivation of the phase relationship
between a VCC and one for conjugated states:
\begin{eqnarray*}
   && \cgc{p_1,q_1;k_1,l_1,m_1}{p_2,q_2;k_2,l_2,m_2}{p,q;k,l,m [j]}
    = \ph{\gamma+2 \sigma +j} \\ && \cdot
   \cgc{q_1,p_1;\tilde{k}_1,\tilde{l}_1,\tilde{m}_1}
   {q_2,p_2;\tilde{k}_2,\tilde{l}_2,\tilde{m}_2}
   {q,p;\tilde{k},\tilde{l},\tilde{m} [j]}.
\end{eqnarray*}

It is tedious to justify the relationship between a VCC and one in
which the final state is interchanged with either of the first two.
Even in the case of SU2, the corresponding formula is often justified as being
a consequence of analytic expressions for the VCC's, rather than proven
from the properties of the SU2 generators.  In the SU3 case,
it is straightforward to show that such permutations leave the
$j$ degeneracy lable unchanged; inspection of the ladder operators
show they transform among one another; the sign convention fixes the
signs of all coefficients; and a numerical multiplication factor is necessary
to preserve normalization.
The resulting relation is
\begin{eqnarray*}
 &&     \cgc{p_1,q_1;k_1,l_1,m_1}{p_2,q_2;k_2,l_2,m_2}{p,q;k,l,m [j]} =
\ph{\gamma+2 \sigma +m_2+j} \sqrt{\frac{(p+1)(q+1)(p+q+2)}
{(p_1+1)(q_1+1)(p_1+q_1+2)}} \\
&& \, \cdot \cgc{q,p;\tilde{k},\tilde{l},\tilde{m}}{p_2,q_2;k_2,l_2,m_2}
{q_1,p_2;\tilde{k}_1,\tilde{l}_1,\tilde{m}_1 [j]}.
\end{eqnarray*}

\medskip\medskip
{\flushleft {\bf VI. Conclusion}}
\medskip

In the previous sections, a presentation has been made of a technique
for derivation of a consistent set of vector coupling coefficients for SU3.
They are discussed in a fashion as near as possible to the standard language
of SU2 as it is used in angular momentum theory.  These VCC's have
symmetry properties under state interchange which parallels the familiar
relations of SU2, and should thus be immediately useful for particle state
classification uses and should lead to 6-j and 9-j recoupling coefficients
with familiar properties of symmetry which should also simplify {\em their}
evaluation.  The algorithm described above is easily implemented for computing,
and such a program has been written in ANSI standard C, and will be made
available to interested users.  A quite similar code, which generates
the SU3 VCC's using the canonical resolution scheme, has been put
in the public domain by the author \nextref{\htw}.  Only minor
modifications of this code are necessary to implement the scheme of this
paper.

The recursion relations for the VCC's for states of highest weight, mentioned
earlier, suggest a much more efficient computer procedure for evaluation
of VCC's than the existing one is possible.  Implementation
of this is presently underway.  The increased computer efficiency which
this scheme will afford will make possible quite prompt evaluation of SU3
6-j and 9-j symbols.  This efficiency provides another commendation of
this scheme in comparison to many previous degeneracy resolution procedures,
which require diagonalization of quite large matrices to determine the VCC's
\nextref{\fn}.

\medskip\medskip
{\flushleft {\bf Acknowledgements}}
\medskip

The author would like to acknowledge the useful guidance received during
several conversations and correspondences with L. C. Biedenharn, and express
appreciation for this help.  Thanks are also due to M. Danos for careful
reading of an early draft of this work, and many useful suggestions.
This work was partially supported by a research
grant from Washington and Lee University.

 \clearpage

\medskip\medskip
\centerline{\bf REFERENCES}
\medskip

\begin{description}
\item{[1]} 
	Variously referred to, as well, as Wigner coefficients,
	Wigner-Clebsch-Gordan coefficients, or simply Clebsch Gordan coefficients.
\item{[2]} 
	L. C. Biedenharn and H. Van Dam, {\em Quantum Theory of Angular Momentum},
	(Academic Press, New York, 1965).
\item{[3]} 
	A. R. Edmonds, {\em Angular Momentum in Quantum Mechanics},
	(Princeton University Press, Princeton, NJ, 1975).
\item{[4]} 
	M. E. Rose, {\em Elementary Theory of Angular Momentum},
	(John Wiley \& Sons, New York, 1957).
\item{[5]} 
	M. Gell Mann, "Symmetries of Baryons and Mesons," Physical Review
	{\bf 125}, 1067-1084 (1962);
	Y. Ne'eman, Nuclear Physics {\bf 26}, 222, (1961).
\item{[6]} 
	J. P. Elliot, Proceedings of the Royal Society {\bf A245}, 128, (1958).
\item{[7]} 
	Consideration here is restricted to the decomposition
	$SU3 \supset SU2 \otimes U1$.
\item{[8]} 
	See, for example M. F. O'Reilly, Journal of Mathematical Physics, {\bf 23},
	2022, (1982).
\item{[9]} 
	J. J. deSwart, "The Octet Model and its Clebsch-Gordan Coefficients,"
	Rev. Mod. Phys. {\bf 35}, 916-939, (1963).
\item{[10]} 
	L. C. Biedenharn, "The Generalisation of the Wigner-Racah Angular
	Momentum Calculus," Phy. Lett. {\bf 3}, 691-700, (1963).
\item{[11]} 
	L. C. Biedenharn, A. Giovannini, J. D. Louck, "Canonical Definition
	of Wigner Coefficients in U(N)," J. Math. Phys. {\bf 8},
	691-700, (1967).
\item{[12]} 
	J. A. Alcaras, L. C. Biedenharn, K. T. Hecht, G. Neely,
	"On the 27-Plet Unitary Operator," Annals of Phys. {\bf 60},
	85-147, (1970).
\item{[13]} 
	L. C. Biedenharn, M. A. Lohe, J. D. Louck, "On the Denominator
	Function for Canonical SU(3) Tensor Operators," J. Math. Phys.,
	{\bf 26}, 1458-1492, (1985).
\item{[14]} 
	J. Alisauskas, "Biorthogonal and Orthonormal SU(3) Clebsch-Gordan
	Coefficients," J. Math. Phys. {\bf33}, 1983-2004,(1992).
\item{[15]} 
	J. D. Louck, L. C. Biedenharn, M. A. Lohe, "On the Denominator
	Function for Canonical SU(3) Operators II," J. Math. Phys.
	{\bf 29}, 1106-1117, (1988).
\item{[16]} 
	L. C. Biedenharn, J. D. Louck, Commun. Math. Physics {\bf 8},
	89-131, (1968).
\item{[17]} 
	J. P. Draayer, Y. Akiyama, "A Pattern Calculus for Tensor Operators
	in the Unitary Groups," J. Math. Phys. {\bf 14},
	1904-1912, (1973).
\item{[18]} 
	Y. Akiyama, J. P. Draayer, "A User's Guide to Fortran Programs
	for Wigner and Racah Coefficients of SU(3),"Computer Physics
	Communications {\bf 5}, 405-415, (1973).
\item{[19]} 
	J. P. Draayer, Y. Leschber, S. C. Park, R. Lopez, "Representations
	of U(3) in U(N)," Computer Physics Communications {\bf 56}, 279-290,
	(1989).
\item{[20]} 
	S. C. Park, J. P. Draayer, Computer Physics Communications {\bf 55},
	189-204, (1989).
\item{[21]} 
	R. LeBlanc, D. J. Rowe, "Balanced Binary Tree Code for Scientific
	Applications," J. Phys. A, {\bf 19}, 2913-2933, (1986).
\item{[22]} 
	D. Braunschweig, "Reduced SU(3) Coefficients of Fractional
	Parentage," Computer Physics Communications {\bf 14}, 109-112, (1978).
\item{[23]} 
	J.-R. Derome, W. T. Sharp, "Racah Algebra for an Arbitrary Group,"
	J. Math. Phys. {\bf 6}, 1584-1590, (1965).
\item{[24]} 
	J.-R. Derome, Symmetry Properties of the 3-j Symbols for an Arbitrary
	Group," J. Math. Phys. {\bf 7}, 612-615, (1966).
\item{[25]} 
	J.-R. Derome, "Symmetry Properties of the 3-j Symbols for SU(3),"
	J. Math. Phys. {\bf 8}, 714-716, (1967).
\item{[26]} 
	Z. Pluhar, Yu. F. Smirnov, V. N. Tolstoy, "Clebsch-Gordan Coefficients
	of SU(3) with Simple Symmetry Properties," J. Phys. A {\bf 19},
	21-28, (1986).
\item{[27]} 
	A. Messiah, {\em Quantum Mechanics, Vol. II} (North-Holland, Amsterdam,
	1962).
\item{[28]} 
	S. Gasiorowicz, S. L. Glashow, {\em Advances in Theoretical Physics, Vol.
	II} (Academic Press, New York, 1966.)
\item{[29]} 
	As, for example, in reference 15.
\item{[30]} 
	H. T. Williams, preprint, June 1993, submitted to Computers in Physics.
\end{description}

\end{document}